\documentclass[prl,twocolumn]{revtex4}
\usepackage{graphicx}
\usepackage[]{epsfig}
\usepackage{times}
\begin{document}
\title{Excited-state spectroscopy on a quantum dot side-coupled to a quantum wire}
\author{T. Otsuka}
\email[]{t-otsuka@issp.u-tokyo.ac.jp}
\affiliation{Institute for Solid State Physics, University of Tokyo, 5-1-5
Kashiwanoha, Chiba 277-8581, Japan}
\author{E. Abe}
\affiliation{Institute for Solid State Physics, University of Tokyo, 5-1-5
Kashiwanoha, Chiba 277-8581, Japan}
\author{Y. Iye}
\affiliation{Institute for Solid State Physics, University of Tokyo, 5-1-5
Kashiwanoha, Chiba 277-8581, Japan}
\author{S. Katsumoto}
\affiliation{Institute for Solid State Physics, University of Tokyo, 5-1-5
Kashiwanoha, Chiba 277-8581, Japan}
\date{\today}
\begin{abstract}
We report excited-state spectroscopy on a quantum dot side-coupled to a quantum wire
with accurate energy estimation.
Our method utilizes periodic voltage pulses on the dot, and 
the energy calibration is performed with reference to the bias voltage across the wire.
We demonstrate the observation of the orbital excited state 
and the Zeeman splitting in a single dot.
\end{abstract}

\pacs{}
\maketitle
Semiconductor quantum dots are promising candidates for quantum bits (qubits)~\cite{1998LossPRA}.
In a conventional dot with two leads,
the current through the dot as a function of the source-drain bias and a gate voltage provides the information of the excited states as well as the ground states~\cite{1997KouwenhovenSci}.
However in the application to quantum information processing, the number of
leads is desired to be as small as possible, because the connection to the outside circuits brings in 
quantum decoherence.
Besides the decoherence, a dot with a single-lead has a number of advantages
such as the spatial compactness, the easiness to go down to the few electron 
regime~\cite{2007OtsukaJPSJ}, and so on.

In single-lead dots, the spectroscopic information is usually given
through
the interference (the Fano effect)~\cite{2004KobayashiPRB,2004JohnsonPRL}
or the charge detection~\cite{1993FieldPRL}.
However, these measurements were
limited to the spectroscopy of the ground states.
To overcome the difficulty, a method combining remote charge sensing and pulsed electrostatic gating
was demonstrated for excited-state spectroscopy on a nearly closed quantum dot~\cite{2004ElzermanAPL}.
Now the remaining difficulty is to find out a ``measure'' for the energy,
in other words the conversion factor from the gate voltage to the energy,
without applying finite voltage nor current across the dots.
So far such conversion is performed by changing the effective dot configuration with gate voltages
or by
numerical simulation. The former inevitably causes significant variation
in the electrostatic parameters, while the latter just gives 
approximate estimation.
In order to make precise meaningful spectroscopy, a method to obtain
the reliable conversion factor is indispensable.

In this letter, we show that the conversion factor can be precisely obtained
in a specially designed single-lead dot, namely a quantum dot side-coupled to a quantum wire.
Our method is based on the one by Elzerman {\it et al.}~\cite{2004ElzermanAPL},
but differs in that the bias voltage can be applied across the quantum wire.
When the wire length is much shorter than the mean free path, the bias on the wire creates two quasi-Fermi levels~\cite{1997PothierPRL, 2002FranceshiPRL}.
If the dot can detect the non-equilibrium energy distribution on the gate voltage axis, this gives the conversion factor.
It is demonstrated that the present method can be applied to the measurement of
spin-splittings as well as orbital excited states.

\begin{figure}[b]
\begin{center}\leavevmode
\includegraphics{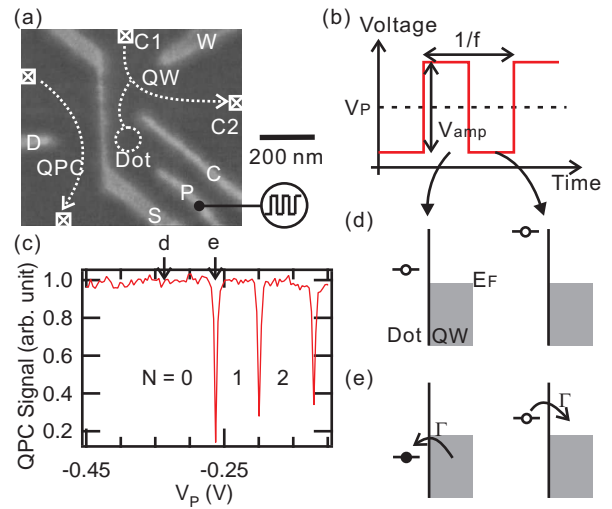}
\caption{(Color online)
(a) Scanning electron micrograph of the device.
(b) Schematic of  pulse train applied on gate P.
(c) QPC signal as a function of $V_{\rm P}$.
The dips reflect the electron response. $N$ is the number of electrons in the dot.
(d, e) Schematic energy diagram for cases without any electron response (d) and with response (e).
In (e), an electron moves into/from the dot with tunneling rate $\Gamma $}
\label{Sample}
\end{center}
\end{figure}

Figure \ref{Sample}(a) shows a scanning electron micrograph of our device fabricated on a GaAs/AlGaAs heterostructure
containing a 2DEG 60 nm below the surface.
Applying negative voltages on gates S, P, C and W,
we prepare a quantum dot which couples to a short quantum wire (QW in the figure).
A quantum point contact (QPC), formed between gates S and D,
is used for remote charge sensing~\cite{1993FieldPRL}.
The potential barrier due to gate S is set high enough so that no exchange of electrons occurs between QPC and the dot. 

First we describe the experiments at zero bias voltage.
A train of voltage pulses is applied on the plunger gate P.
The pulse train is a rectangular wave with duty ratio 0.5,
and is characterized by amplitude $V_{\rm amp}$, repetition frequency $f$ and center voltage $V_{\rm P}$ [Fig.~\ref{Sample}(b)].
The synchronized current oscillation in QPC is lock-in detected.

The voltage pulses applied on gate P shift the energy levels in the dot up and down.
If an energy level of the dot is always above the Fermi level of QW $E_{\rm F}$,
the level is always empty and cannot exchange electrons with QW [Fig.~\ref{Sample}(d)].
The signal in this case reflects the direct capacitive coupling between gate P and QPC.
When $V_{\rm P}$ is shifted and $E_{\rm F}$ gets into the region where
the level is swinging, and if $f$ is low enough,
the dot captures and emits an electron
in phase with the pulse [Fig. \ref{Sample}(e)].
The captured electron compensates the effect of the pulsed voltage on gate P, thus the current oscillation through QPC becomes smaller.

Figure~\ref{Sample}(c) shows a typical signal of QPC as a function of $V_{\rm P}$ (the coupling gate voltage $V_{\rm C}=-0.45$ V, $V_{\rm amp} = 4.5$~mV and $f = 474$~Hz),
in which three dips are observed.
Here, we operate in the regime where tunneling time between the wire and the dot
$1/\Gamma $ is sufficiently shorter than the pulse duration $\tau =1/(2f)$ and the charge-up of the dot
completely follows the pulses.
For $V_{\rm P}<-0.26$ V, dips are no longer observed, indicating full depletion of electrons in the dot.

\begin{figure}[]
\begin{center}\leavevmode
\includegraphics{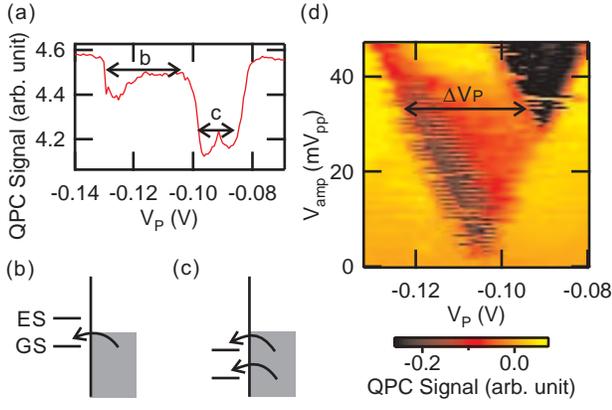}
\caption{(Color online)
(a) QPC signal as a function of $V_{\rm P}$. $V_{\rm C}=-0.52$ V, $V_{\rm amp} = 42$~mV and $f = 735$~Hz.
(b, c) Schematic energy diagram when the total voltage on gate P is $V_{\rm P}+V_{\rm amp}/2$.
In (b), only the ground state is accessible. In (c), an excited state is also accessible, and
the effective tunneling rate becomes larger. (d) QPC signal as a function of $V_{\rm P}$ and $V_{\rm amp}$.}
\label{Orbital}
\end{center}
\end{figure}

The following experiments are carried out around a dip between zero- and one-electron states.
$f$ and $V_{\rm C}$ are optimized for making $\tau $ comparable with $1/\Gamma $.
The QPC signal at $V_{\rm C}=-0.52$ V, $V_{\rm amp}=42$ mV and $f=735$ Hz is shown in Fig. \ref{Orbital}(a).
The electron response produces a hollow in $-0.13 < V_{\rm P} < -0.08$~V.
This corresponds to an ``energy window'', within which the electrons can go to and from the dot.
The hollow is deeper in the less negative side (region c).

This two-step structure is a hallmark of the presence of the excited state and
interpreted as follows.
Because $\tau \sim 1/\Gamma $, the charging of the dot stochastically follows the pulses.
In region b, movement of electrons occurs only via the ground state [Fig.~\ref{Orbital}(b)].
When the total voltage on gate P is $V_{\rm P}+V_{\rm amp}/2$, only the ground state is below 
$E_{\rm F}$ and can accept an electron from QW with tunneling rate $\Gamma _{\rm g}$.
On the other hand, in region c, an electron can flow into both the ground and excited states [Fig.~\ref{Orbital}(c)].
This makes effective tunneling rate $\Gamma_{\rm eff} = \Gamma _{\rm g}+\Gamma _{\rm e}$ larger than  $\Gamma_{\rm g}$ and the signals take smaller values, where $\Gamma _{\rm e}$ is the tunneling rate of the excited state.
Since the measurement is performed in zero-magnetic field and one-electron state,
the excited state is attributed to orbital one.
Note that simultaneous occupation with two electrons is prohibited by the
charging energy and the change in the QPC signal is solely due to that in $\Gamma_{\rm eff}$.

Figure~\ref{Orbital}(d) shows the QPC signal as a function of $V_{\rm amp}$ and $V_{\rm P}$.
For clarity, we subtracted the background that comes from the direct coupling to gate P.
As expected, the width of the hollow region agrees with $V_{\rm amp}$.
The deeper hollow region appears around $V_{\rm amp} = 30$~mV.
When the excited state is accessible,
the width of the shallow region is constant against $V_{\rm amp}$, and denoted as $\Delta V_{\rm P}$ in the figure.
$\Delta V_{\rm P}$ reflects the energy difference between the ground and the first excited orbital states $\Delta \epsilon $.
The two quantities are connected by a conversion factor $\alpha $ ($\Delta \epsilon = \alpha \Delta V_{\rm P}$).

\begin{figure}[]
\begin{center}\leavevmode
\includegraphics{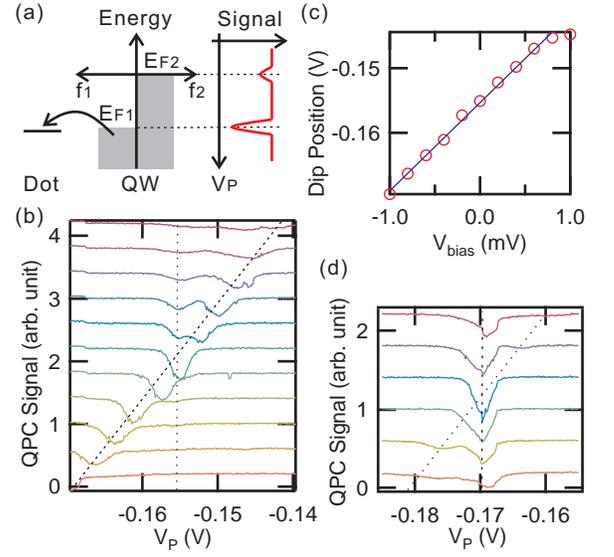}
\caption{(Color online)
(a) Schematic energy diagram and the resulting QPC signal when $V_{\rm bias}$ is applied across QW. Non-equilibrium energy distribution is formed and produces two dips in the signal.
(b) Shift of the dip position with $V_{\rm bias}$ applied on contact C1. $V_{\rm bias}$ is changed from $-$1~mV (bottom) to 1~mV (top). Traces are vertically offset by 0.5.
Dotted lines are guides for the movement of dips.
(c) Dip position as a function of $V_{\rm bias}$. The solid line is a fit.
(d) Shift of the dip position when $V_{\rm bias}$ is applied on contact C2.}
\label{Bias}
\end{center}
\end{figure}

Next, to estimate the value of $\alpha$, we apply finite bias voltage $V_{\rm bias}$ on contact C1, and contact C2 is grounded.
Since the mean free path of the electrons is about $3~\mu$m, and is much longer than QW,
the energy relaxation in QW is negligible.
This produces non-equilibrium energy distribution inside QW, as illustrated in Fig.~\ref{Bias}(a)~\cite{1997PothierPRL, 2002FranceshiPRL}. There are two quasi-Fermi levels $E_{\rm F1}$ and $E_{\rm F2}$, which correspond to 
C1 and C2 respectively.
Now the dot couples to these two kinds of electrons and produces
signals for exchanging the electrons with them.
The condition of $V_{\rm amp}$ 
is adjusted so that only the ground state may be in the energy window.
Hence, two dips due to $E_{\rm F1}$ and $E_{\rm F2}$ are expected 
in the QPC signal.

Figure~\ref{Bias}(b) shows the QPC signal with changing $V_{\rm bias}$ from -1 mV (bottom) to 1 mV (top).
We in fact observe two types of dips,
one of which shifts the position with $V_{\rm bias}$, while the other is fixed around $V_{\rm P}=-0.156$~V.
The former corresponds to $E_{\rm F1}$.
The relation between the dip position and $V_{\rm bias}$ shows excellent linearity.
We obtain $\alpha$ by linear fitting [Fig.~\ref{Bias}(c)].
Here, the contribution from series resistance, which is estimated from the resistance when we remove the confinement of QW, has also been subtracted.
With this procedure, we obtain the value $\alpha = 0.075$ eV/V.
Then the energy difference between the ground and excited orbitals $\Delta \epsilon$ is given 
as 2.3 meV.
We would like to emphasize that this calibration can be performed without changing
any relevant parameter of the side-coupled system.

In Fig.~\ref{Bias}(b), the shifting dips have higher visibility, reflecting the difference in the coupling between the dot and the two quasi-Fermi levels.
In our device, the entrance of the dot is directed to C1 [Fig. 1(a)].
This asymmetry would make stronger coupling between C1 and the dot.
This is further confirmed by grounding C1 and sweeping the voltage of C2.
As shown in Fig.~\ref{Bias}(d), the clearer dips for C1 are fixed and the others shift with the bias on C2.
These also support the picture that the two quasi-Fermi levels are formed and the dot couples to them.

\begin{figure}[]
\begin{center}\leavevmode
\includegraphics{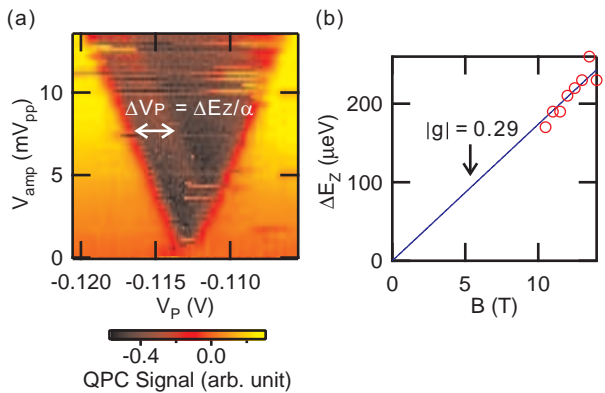}
\caption{(Color online) (a) QPC signal as a function of $V_{\rm P}$ and $V_{\rm amp}$ at 14 T.
A two-step hollow structure is observed. The width of the shallow region corresponds to the Zeeman splitting.
(b) Zeeman splitting as a function of magnetic fields. A linear fit gives $|g|=0.29$.}
\label{Zeeman}
\end{center}
\end{figure}

For further checking the validity of our method, we now apply it to the Zeeman splitting.
Magnetic fields $B$ lift the spin degeneracy and produce the Zeeman splitting $\Delta E_{\rm Z}=g\mu _{\rm B}B$,
where $g$ is the g-factor and $\mu _{\rm B}$ is the Bohr magneton. 
It has been reported that the g-factor in heterostructure is smaller than the bulk value of $g = -0.44$ in GaAs~\cite{1988DobersPRB,2003PotokPRL,2003HansonPRL,2005BeverenNJP}.
We apply in-plane magnetic fields in order to minimize the magnetic effect on orbital states of the dot.
The QPC signal as a function of $V_{\rm P}$ and $V_{\rm amp}$ at 14 T is shown in Fig. \ref{Zeeman}(a).
Here, the range of $V_{\rm amp}$ is limited so that the orbital excited state is not visible.
The background is again subtracted to improve visibility.
The two-step hollow structure similar to Fig. \ref{Orbital}(d) is observed.
In this case, the excited state is due to lifting of the Kramers degeneracy by magnetic field
and the width of the shallow region reflects $\Delta E_{\rm Z}$.
By calibrating the conversion factor at each field,
we obtain $\Delta E_{\rm Z}$ as a function of magnetic field, which is shown in Fig.~\ref{Zeeman}(b).
Though the signal-to-noise ratio was not high enough to resolve the Zeeman splitting at lower fields, the data at high fields align along a line pointing the origin as shown in Fig.~\ref{Zeeman}(b).
The gradient gives $|g|=0.29$ in agreement
with previous reports on transport measurements~\cite{2003PotokPRL,2003HansonPRL,2005BeverenNJP},
certifying the validity and the effectiveness of the present method.

In this work, we demonstrate excited-state spectroscopy on a side-coupled quantum dot with accurate energy estimation
by utilizing voltage pulse on the dot and energy calibration with bias voltage over the wire.
The orbital excited state 
as well as the Zeeman splitting are detected with this method and
the obtained g-factor in the dot is in good agreement with the measurements so far reported. 

We thank Y. Hashimoto and S. Tamiya for technical supports.
This work is supported by Grant-in-Aid for Scientific Research and Special Coordination Funds for Promoting Science and Technology.

\end{document}